\documentclass[prl,twocolumn,showpacs,superscriptaddress]{revtex4}

\usepackage{amsmath,amssymb}
\usepackage{verbatim}
\usepackage{graphicx}
\usepackage{hyperref}
\usepackage{color} 

\DeclareFontFamily{OT1}{rsfs}{}
\DeclareFontShape{OT1}{rsfs}{m}{n}{ <-7> rsfs5 <7-10> rsfs7 <10->rsfs10}{} 
\DeclareMathAlphabet{\mycal}{OT1}{rsfs}{m}{n}

\newcommand{\eq}[2]{\begin{equation} #1 \label{#2} \end{equation}}

\newcommand{\al}{\alpha}
\newcommand{\ga}{\gamma}
\newcommand{\de}{\delta}

\newcommand{\si}{\sigma}

\newcommand{\Om}{\Omega}

\DeclareMathOperator{\extdm}{d}
\newcommand{\extd}{\extdm \!}

\newcommand{\ts}[1]{\textrm{\tiny #1}}
\newcommand{\ms}[1]{\textrm{\tiny $#1$}}
\newcommand{\LO}{\ms{(0)}}
\newcommand{\FO}{\ms{(1)}}
\newcommand{\SO}{\ms{(2)}}
\newcommand{\TO}{\ms{(3)}}
\newcommand{\NO}{\ms{(n)}}

\newcommand{\killi}{k} 

\begin{document}

\title{Conformal gravity holography in four dimensions}

\author{Daniel Grumiller}
\email{grumil@hep.itp.tuwien.ac.at}
\affiliation{Institute for Theoretical Physics, Vienna University of Technology, Wiedner Hauptstrasse 8--10/136, A-1040 Vienna, Austria}

\author{Maria Irakleidou}
\email{irakleidou@hep.itp.tuwien.ac.at}
\affiliation{Institute for Theoretical Physics, Vienna University of Technology, Wiedner Hauptstrasse 8--10/136, A-1040 Vienna, Austria}

\author{Iva Lovrekovic}
\email{lovrekovic@hep.itp.tuwien.ac.at}
\affiliation{Institute for Theoretical Physics, Vienna University of Technology, Wiedner Hauptstrasse 8--10/136, A-1040 Vienna, Austria}

\author{Robert McNees}
\email{rmcnees@luc.edu}
\affiliation{Loyola University Chicago, Department of Physics, Chicago, IL 60660}

\date{\today}

\preprint{TUW--13--xx}

\begin{abstract} 
We formulate four-dimensional conformal gravity with (Anti-)de~Sitter boundary conditions that are weaker than Starobinsky boundary conditions, allowing for an asymptotically subleading Rindler term concurrent with a recent model for gravity at large distances. 
We prove the consistency of the variational principle and derive the holographic response functions. One of them is the conformal gravity version of the Brown--York stress tensor, the other is a `partially massless response'. The on-shell action and response functions are finite and do not require holographic renormalization.
Finally, we discuss phenomenologically interesting examples, including the most general spherically symmetric solutions and rotating black hole solutions with partially massless hair. 
\end{abstract}


\maketitle

Conformal gravity (CG) in four dimensions is a recurrent topic in theoretical physics, as it provides a possible resolution to some of the problematic issues with Einstein gravity, the established theory of the gravitational interaction, though it usually introduces new ones. 

For instance, like other higher-derivative theories, CG is renormalizable \cite{Stelle:1976gc,Adler:1982ri}, but has ghosts \footnote{%
Generic higher-derivative theories are usually assumed to suffer from ghosts, but it has been conjectured that CG may admit an alternative quantization that preserves unitarity \cite{Bender:2007wu,Mannheim:2011ds} (see \cite{Smilga:2008pr,Chen:2012au} for some criticism). Resolving this question is beyond the scope of this paper. Instead, we focus on the boundary conditions and variational formulation of the classical theory, and on establishing the framework for a possible holographic dual.
}, whereas Einstein gravity is ghost-free, but 2-loop non-renormalizable \cite{Goroff:1986th}. See \cite{Julve:1978xn,Fradkin:1981iu,Tomboulis:1983sw,Boulware:1983td} for important early work on CG. Later, CG was studied phenomenologically by Mannheim in an attempt to explain galactic rotation curves without dark matter \cite{Mannheim:1988dj,Mannheim:2005bfa,Mannheim:2010ti,Mannheim:2011ds} and emerges theoretically from twistor string theory \cite{Berkovits:2004jj} or as a counter term in the Anti-de~Sitter/Conformal Field Theory (AdS/CFT) correspondence \cite{Liu:1998bu,Balasubramanian:2000pq}. More recently, 't~Hooft has studied CG in a quantum gravity context \cite{tHooft:2011aa} and Maldacena has shown how Einstein gravity can emerge from CG upon imposing suitable boundary conditions that eliminate ghosts \cite{Maldacena:2011mk}. 

Physical theories in general require boundary conditions as part of their definition. In many cases ``natural'' boundary conditions --- the rapid fall-off of all fields near a boundary or in an asymptotic region --- are the appropriate choice, but this is not the case in gravitational theories, since the metric should be non-zero. A prime example is gravity in AdS, where the boundary conditions define the behavior of the dual field theory that lives on the conformal boundary of spacetime. Gravity in de~Sitter (dS) requires similar boundary conditions; they were provided for four-dimensional Einstein gravity by Starobinsky \cite{Starobinsky:1982mr}. (See also \cite{Anninos:2010zf,Anninos:2011jp} for a more recent discussion of future boundary conditions and conserved charges in dS.) These boundary conditions played a crucial role in Maldacena's reduction from CG to solutions of Einstein gravity \cite{Maldacena:2011mk}.
  
It is, however, not clear that the Starobinsky boundary conditions are the most general or phenomenologically interesting ones for CG. 
Experience with three-dimensional (3D) CG \cite{Afshar:2011yh} 
suggests that a weaker set of boundary conditions should be possible also in four dimensions. Finding such boundary conditions is interesting for purely theoretical reasons and also phenomenologically. Indeed, the CG analogue of the Schwarzschild solution, the spherically symmetric Mannheim--Kazanas--Riegert (MKR) solution \cite{Riegert:1984zz,Mannheim:1988dj}, does not obey the Starobinsky boundary conditions. A related motivation is to investigate whether it is true that CG provides an example of a theory that allows a non-trivial Rindler term, as suggested in the discussion of an effective model for gravity at large distances \cite{Grumiller:2010bz}. The difficulty does not lie in showing that the CG equations of motion (EOM) permit a Rindler term (they do), but in determining a set of boundary conditions consistent with such a term.

The main purpose of our Letter is to establish the consistency of a set of (A)dS boundary conditions for CG, weaker than the ones proposed by Starobinsky, that are compatible with the existence of an asymptotic Rindler term, the MKR solution and other solutions with a condensate of partially massless gravitons. 

Before starting we review the most salient features of CG. A distinguishing property of CG is that the theory depends only on (Lorentz-)angles but not on distances. This means that the theory is invariant under local Weyl rescalings of the metric,
\eq{
g_{\mu\nu} \to\tilde g_{\mu\nu} = e^{2\omega}\,g_{\mu\nu}\,,
}{eq:CG1}
where the Weyl factor $\omega$ is allowed to depend on the coordinates. 
The bulk action of CG
\eq{
I_{\textrm{\tiny CG}} = \alpha_{\textrm{\tiny CG}} \int\extd^4x\sqrt{|g|} \, g_{\al\mu}g^{\beta\nu}g^{\ga\lambda}g^{\de\tau}\,C^\al{}_{\beta\ga\de} C^\mu{}_{\nu\lambda\tau}
}{eq:CG2}
is manifestly invariant under Weyl rescalings \eqref{eq:CG1}, since the Weyl tensor $C^\al{}_{\beta\ga\de}$ is Weyl-invariant, and the Weyl factor coming from the square root of the determinant of the metric is precisely cancelled by the Weyl factor coming from the metric terms. The dimensionless coupling constant $\alpha_{\textrm{\tiny CG}}$ is the only free parameter in the CG action. In most of what follows we set $\alpha_{\textrm{\tiny CG}}=1$ to reduce clutter.
The EOM of CG are fourth order and require the vanishing of the Bach tensor,
\eq{
\big(\nabla^\delta\nabla_\ga + \frac12\,R^\delta{}_\ga\big)\, C^\ga{}_{\al\de\beta} = 0\,.
}{eq:CG3}
There are two especially simple classes of solutions to the EOM: conformally flat metrics ($C^\ga{}_{\al\de\beta}=0$) and Einstein metrics ($R_{\al\beta}\propto g_{\al\beta}$) both have vanishing Bach tensor. 
Therefore, solutions of Einstein gravity are a subset of the broader class of solutions of CG. 

The most general spherically symmetric solution of CG is given by the line-element \cite{Riegert:1984zz}
\eq{
\extd s^2 = -\killi(r)\extd t^2+\frac{\extd r^2}{\killi(r)}+r^2\,\extd\Om^2_{S^2}
}{eq:CG4}
where $\extd\Om^2_{S^2}$ is the line-element of the round 2-sphere and 
\eq{
\killi(r)=\sqrt{1-12aM}-\frac{2M}{r}-\Lambda r^2+2ar 
}{eq:CG5}
For $a=0$ the solution reduces to Schwarzschild-(A)dS.
It is noteworthy that for $aM\ll 1$ the solution \eqref{eq:CG4}, \eqref{eq:CG5} corresponds to the one presented in \cite{Grumiller:2010bz}, derived from an effective model for gravity at large distances. Phenomenologically relevant numbers (in Planck units) are $\Lambda\approx 10^{-123}$, $a\approx 10^{-61}$, $M\approx 10^{38} M_\odot$, where $M_\odot=1$ for the Sun, so that indeed $aM\approx 10^{-23}M_\odot \ll 1$ for all black holes or galaxies in our Universe.

We propose now boundary conditions that admit the MKR solution \eqref{eq:CG4}, \eqref{eq:CG5}. This requires the introduction of a length scale $\ell$, which in Einstein gravity would be related to the cosmological constant as $\Lambda = 3\sigma/\ell^2$ (with $\sigma=-1$ for AdS and $\sigma=+1$ for dS). Then our asymptotic ($0<\rho\ll \ell$) line-element reads
\eq{
\extd s^2 =  \frac{\ell^2}{\rho^2}\,\big(- \sigma \extd\rho^2 + \gamma_{ij}\,\extd x^i\extd x^j\big) ~.
}{eq:CG6}
For simplicity we partially fix the gauge and use Gaussian coordinates. 
Close to the conformal boundary at $\rho=0$ the 3D metric has the following asymptotic expansion:
\eq{
\gamma_{ij} = \gamma_{ij}^{\LO} + \frac{\rho}{\ell}\,\gamma_{ij}^{\FO} + \frac{\rho^2}{\ell^2}\,\gamma_{ij}^{\SO} + \frac{\rho^3}{\ell^3}\, \gamma_{ij}^{\TO} + \dots
}{eq:CG7}
The boundary metric $\ga^{\LO}$ is required to be invertible. All the coefficient matrices $\ga^{\NO}$ are allowed to depend on the boundary coordinates $x^i$, but not on the ``holographic'' coordinate $\rho$. 

As part of the specification of our boundary conditions we fix the leading and first-order terms in \eqref{eq:CG7} on $\partial{\cal M}$ up to a local Weyl rescaling
\eq{
\delta\gamma_{ij}^{\LO}|_{\partial{\cal M}} = 2\,\lambda\,\ga^{\LO}_{ij} \qquad  \delta\gamma_{ij}^{\FO}|_{\partial{\cal M}} = \lambda\,\ga^{\FO}_{ij}\,
}{eq:CG8}
where $\lambda$ is a regular function on $\partial {\cal M}$,
while the subleading terms at second and higher order are allowed to vary freely, $\delta\gamma^{\NO}|_{\partial{\cal M}} \neq 0$ for $n\geq 2$. An essential difference to the Starobinsky boundary conditions is the presence of a subleading term $\ga_{ij}^{\FO}$. This term is absent in \cite{Starobinsky:1982mr} because the EOM for Einstein gravity force it to vanish. By contrast, the EOM \eqref{eq:CG3} do not give any conditions on $\ga_{ij}^{\FO}$, analogous to 3D CG \cite{Afshar:2011yh}. 

 
To check the consistency of the boundary conditions \eqref{eq:CG6}-\eqref{eq:CG8} we consider first the on-shell action and then the variational principle. On general grounds, one might expect the bulk action \eqref{eq:CG2} to be supplemented by two kinds of boundary terms: a ``Gibbons--Hawking--York'' boundary term \cite{York:1972sj, Gibbons:1977ue} that produces the desired boundary value problem (for instance, a Dirichlet boundary value problem), and holographic counterterms \cite{Henningson:1998ey,Balasubramanian:1999re,Emparan:1999pm,Kraus:1999di,deHaro:2000xn,Papadimitriou:2005ii} that guarantee that the action is stationary for all variations that preserve our boundary conditions.  
 
We claim that no such boundary terms are required for CG, so that the full action is just the bulk action \eqref{eq:CG2}
\eq{
\Gamma_{\textrm{\tiny CG}} = I_{\textrm{\tiny CG}} = \int_{\cal M}\!\!\extd^4x\sqrt{|g|}\,C^{\lambda}{}_{\mu\sigma\nu}\,C_{\lambda}{}^{\mu\sigma\nu}\,.
}{eq:CG11}
The first piece of evidence that no counterterms might be needed comes from the calculation of the on-shell action. It is straightforward to show that the on-shell action for any metric behaving like \eqref{eq:CG6}, \eqref{eq:CG7}, evaluated on a compact region $\rho_c \leq \rho$, remains finite as $\rho_c \to 0$. 
A related piece of evidence was provided in \cite{Lu:2012xu}, where it was shown that the free energy derived from the on-shell action \eqref{eq:CG11} is consistent with the Arnowitt--Deser--Misner mass and Wald's definition of the entropy \cite{Wald:1993nt}. The fact that the on-shell action yields the correct free energy suggests that any boundary terms added to the action \eqref{eq:CG11} should vanish on-shell. The simplest possibility is that these terms are identically zero \footnote{%
We stress that vanishing boundary terms are not obviously expected for CG. For example, naively extrapolating the result for boundary terms (see appendix A of \cite{Johansson:2012fs}) in L\"u--Pope massive gravity \cite{Lu:2011zk} would indicate a non-zero result for these terms in CG. See also \cite{Hohm:2010jc}.
}.

A more stringent check of our claim is obtained by proving the consistency of the variational principle and the finiteness of the holographic response functions. To this end we first rewrite the Weyl-squared action \eqref{eq:CG11} as
\begin{multline}
\Gamma_{\textrm{\tiny CG}} = \int_{\cal M}\!\!\extd^4x\sqrt{|g|}\,\big(2\,R^{\mu\nu} R_{\mu\nu} - \tfrac{2}{3}\,R^2\big) \\
+32\pi^2\chi({\cal M}) + \int_{\partial{\cal M}}\!\!\!\!\extd^3x\sqrt{|\ga|}\,\big(-8\,\sigma\,{\cal G}^{ij}K_{ij} \\
+\tfrac{4}{3}\,K^3-4 \,KK^{ij}K_{ij}+ \tfrac{8}{3}\, K^{ij}K_j{}^kK_{ki}\big)
\label{eq:CG12}
\end{multline}
The action has been separated into a topological part -- the Euler characteristic $\chi({\cal M})$ --  and a Ricci-squared action, with the boundary terms in the last two lines canceling similar terms that appear in the Euler characteristic for spacetimes with (conformal) boundary; see \cite{Myers:1987yn}. Here and in all subsequent expressions, calligraphic letters indicate quantities intrinsic to the 3D surface $\partial {\cal M}$. Thus, ${\cal G}^{ij}$ is the 3D Einstein tensor for the metric $\ga_{ij}$. The extrinsic curvature is defined as $K_{ij} = -\tfrac{\sigma}{2} \pounds_{n} \ga_{ij}$, where $\pounds_{n}$ is the Lie derivative along the outward- or future-pointing unit vector $n^{\mu}$ normal to $\partial {\cal M}$.

In this formulation, the first variation of the action is given by
\eq{
\delta\Gamma_{\textrm{\tiny CG}} = \textrm{EOM} + \int_{\partial{\cal M}}\!\!\!\!\extd^3x\sqrt{|\ga|}\,\big(\pi^{ij}\,\de\gamma_{ij} + \Pi^{ij}\,\de K_{ij}\big)~.
}{eq:CG13}
The momentum $\pi^{ij}$ reads
\begin{multline} 
\pi^{ij} = \tfrac{\sigma}{4}(\ga^{ij} K^{kl}-\ga^{kl} K^{ij}) f_{kl} + \tfrac{\sigma}{4} f^{\rho}{}_{\rho} (\ga^{ij} K - K^{ij}) \\
 - \tfrac12 \ga^{ij} {\cal D}_{k}(n_{\rho} f^{k \rho}) + \tfrac{1}{2} {\cal D}^{i}(n_{\rho} f^{\rho j}) - \tfrac{1}{4} (\ga^{ik} \ga^{jl} - \ga^{ij} \ga^{kl}) \pounds_{n} f_{kl} \\
 + \sigma\,\big(2 K {\cal R}^{ij} - 4 K^{ik} {\cal R}_{k}{}^{j}  + 2 \ga^{ij}  K_{kl} {\cal R}^{kl} - \ga^{ij} K {\cal R} + 2 {\cal D}^{2} K^{ij}\\ - 4 {\cal D}^{i} {\cal D}_{k} K^{kj} 
+ 2 {\cal D}^{i} {\cal D}^{j} K  + 2 \ga^{ij} ({\cal D}_{k} {\cal D}_{l} K^{kl} - {\cal D}^{k} {\cal D}_{k} K) \big) \\ 
+ \tfrac{2}{3} \ga^{ij} K^{k}{}_{m} K^{lm} K_{kl} - 4 K^{i k} K^{jl} K_{kl} + 2 K^{ij} K^{kl} K_{kl} + \tfrac{1}{3} \ga^{ij} K^3 \\
- 2 K^{ij} K^2 -\ga^{ij} K K^{kl} K_{kl} + 4 K K^{i}{}_{k} K^{jk} 
 + i \leftrightarrow j\,.
\label{eq:CG14}
\end{multline}
The tensor $f_{\mu\nu}$, which appears in a convenient auxiliary field formulation of the action, is proportional to the four-dimensional Schouten tensor, $f_{\mu\nu} = - 4 (R_{\mu\nu} - \tfrac16 g_{\mu\nu} R)$. The momentum  $\Pi^{ij}$ reads  
\begin{multline}
\Pi^{ij} = -8 \,\sigma\, {\cal G}^{ij} - \sigma \,\big(f^{ij} - \ga^{ij}f^k{}_k \big) \\
+ 4\ga^{ij} \big(K^2 - K^{kl}K_{kl}\big) - 8 K K^{ij} + 8K^i{}_k K^{kj} \,.
\label{eq:CG15}
\end{multline} 
It is noteworthy that we allow the boundary metric and the extrinsic curvature to vary independently in \eqref{eq:CG13}. 

Let us check now the variational principle. Evaluating \eqref{eq:CG13} on a compact region $\rho_c \leq \rho$, applying the EOM, and making use of the asymptotic expansion \eqref{eq:CG7} with \eqref{eq:CG14}, \eqref{eq:CG15} yields
\eq{
\delta\Gamma_{\textrm{\tiny CG}}\big|_{\textrm{\tiny EOM}} = \int_{\partial{\cal M}}\!\!\!\!\extd^3x\sqrt{|\ga^{\LO}|}\,\big(\tau^{ij}\,\de\ga^{\LO}_{ij} + P^{ij}\,\de\ga_{ij}^{\FO}\big)\,.
}{eq:CG16}
The tensors $\tau^{ij}$ and $P^{ij}$ are finite as $\rho_c \to 0$. As we will show below, they satisfy the trace conditions ($\psi^{\FO}_{ij}:=\ga^{\FO}_{ij} - \tfrac13\,\ga^{\LO}_{ij}\ga^{\FO k}_k$)
\begin{gather} \label{eq:Trace}
 \ga_{ij}^{\LO} \tau^{ij} + \tfrac12\,\psi^{\FO}_{ij} P^{ij} = 0 \qquad 
 \ga_{ij}^{\LO} P^{ij} = 0 ~,
\end{gather}
so that the first variation of the action vanishes on-shell when the boundary conditions \eqref{eq:CG8} are satisfied. Therefore the action \eqref{eq:CG11} and our proposed boundary conditions constitute a well-defined variational principle.



The quantities $\tau^{ij}$ and $P^{ij}$ appearing in \eqref{eq:CG16} are the holographic response functions conjugate to the sources $\ga^{\LO}_{ij}$ and $\ga^{\FO}_{ij}$, respectively. We evaluate now the first of these functions, which is proportional to the usual 
Brown--York stress tensor. 
It is useful to introduce the electric $E_{ij}$ and magnetic $B_{ijk}$ parts of the Weyl tensor. 
\eq{
E_{ij} = n_\al n^\beta C^\al{}_{i\beta j}\qquad B_{ijk} = n_\al C^\al{}_{ijk}
}{eq:CG20}
For metrics that satisfy the boundary conditions \eqref{eq:CG6}, \eqref{eq:CG7} their asymptotic expansions are given by
\begin{align}
E_{ij} &= 
E_{ij}^{\SO} + \tfrac{\rho}{\ell} E_{ij}^{\TO} + \dots \label{eq:CG21}\\
B_{ijk} &= 
\tfrac{\ell}{\rho}\,B_{ijk}^{\FO} + B_{ijk}^{\SO} + \dots \label{eq:CG22}
\end{align} 
with 
\begin{align}
& B_{ijk}^{\FO} = \tfrac{1}{2\ell}\,\big({\cal D}_j\psi^{\FO}_{ik}-\tfrac12\,\ga_{ij}^{\LO}\,{\cal D}^l\psi^{\FO}_{kl}\big) - j \leftrightarrow k \label{eq:CG24} \\
& E_{ij}^{\SO} =  - \tfrac{1}{2\ell^2} \psi_{ij}^{\SO}  + \tfrac{\sigma}{2}\, \big({\cal R}_{ij}^{\LO} - \tfrac13 \ga_{ij}^{\LO}{\cal R}^{\LO}\big) + \tfrac{1}{8\ell^2} \gamma^{\FO} \psi_{ij}^{\FO} \label{eq:CG23} \\
& E^{\TO}_{ij} = -\tfrac{3}{4\ell^2}\,\psi^{\TO}_{ij} -\tfrac{1}{12\ell^{2}}\,\ga_{ij}^{\LO}\,\psi^{kl}_{\FO} \, \psi_{kl}^{\SO}
-\tfrac{1}{16\ell^{2}}\,\psi^{\FO}_{ij}\,\psi^{\FO}_{kl}\,\psi_{\FO}^{kl} \nonumber \\
& - \tfrac{\si}{12}\,\big(\mathcal{R}^{\LO}\,\psi_{ij}^{\FO}-\ga _{ij}^{\LO}\,\mathcal{R}_{kl}^{\LO}\,\psi^{kl}_{\FO}
+\ga _{ij}^{\LO}\,\mathcal{D}_{l}\,\mathcal{D}_{k}\,\psi^{kl}_{\FO} \nonumber \\
& +\tfrac{3}{2}\,\mathcal{D}_{k}\,\mathcal{D}^{k}\,\psi_{ij}^{\FO}
-3\,\mathcal{D}_{k}\,\mathcal{D}_{i}\,\psi^{\FO k}_{j} 
\big) + \tfrac{1}{24\ell^2} \, E_{ij}^{
\gamma} + i \leftrightarrow j
\label{eq:CG26a} \\
& E_{ij}^{
\gamma}  :=  \ga_{\FO}\,(3\,\psi^{\SO}_{ij}+\tfrac12\,\ga_{ij}^{\LO}\,
\psi_{kl}^{\FO}\,\psi^{kl}_{\FO}-\gamma_{\FO}\,\psi_{ij}^{\FO}) \nonumber \\
&  + 5\,\ga _{\SO}\,\psi_{ij}^{\FO} - \si\ell^2\,(\mathcal{D}_{j}\,\mathcal{D}_{i}\,\ga_{\FO} 
-\tfrac{1}{3}\,\ga _{ij}^{\LO}\,\mathcal{D}^{k}\,\mathcal{D}_{k}\,\ga _{\FO})\,.
\label{eq:CG26b}
\end{align}
In these expressions the coefficient matrices have been split into trace and trace-free parts as
$\ga_{ij}^{\NO} = \tfrac13\,\ga_{ij}^{\LO}\,\ga^{\NO}+\psi_{ij}^{\NO}$.
Then the (
finite) result for $\tau_{ij}$ (as $\rho_c \to 0$) is
\begin{multline}
\tau_{ij} = \sigma \big[\tfrac{2}{\ell}\,(E_{ij}^{\TO}+ \tfrac{1}{3} E_{ij}^{\SO}\ga^{\FO}) -\tfrac4\ell\,E_{ik}^{\SO}\psi^{\FO k}_j
+ \tfrac{1}{\ell}\,\ga_{ij}^{\LO} E_{kl}^{\SO}\psi_{\FO}^{kl}\\ 
+ \tfrac{1}{2\ell^3}\,\psi^{\FO}_{ij}\psi_{kl}^{\FO}\psi_{\FO}^{kl}
- \tfrac{1}{\ell^3}\,\psi_{kl}^{\FO}\,\big(\psi^{\FO k}_i\psi^{\FO l}_j-\tfrac13\,\ga^{\LO}_{ij}\psi^{\FO k}_m\psi_{\FO}^{lm}\big)\big] \\
- 4\,{\cal D}^k B_{ijk}^{\FO} + i\leftrightarrow j\,.
\label{eq:CG17}
\end{multline}

We call the function $P^{ij}$ the ``partially massless response'' (PMR).
This name is justified, since it is sourced by the term $\ga_{ij}^{\FO}$ in the metric. The latter, when plugged into the linearized CG EOM around an (A)dS background, exhibits partial masslessness in the sense of Deser, Nepomechie and Waldron \cite{Deser:1983mm,Deser:2001pe}. This is expected from the corresponding behavior in 3D \cite{Afshar:2011yh} 
and also on general grounds, since the Weyl invariance \eqref{eq:CG1} is nothing but the non-linear completion of the gauge enhancement at the linearized level due to partial masslessness; see for instance the recent discussion in \cite{Deser:2013bs,Deser:2013uy}. (Note that such a non-perturbative completion of partial masslessness is not generic to higher derivative theories \cite{Deser:2012ci}.) Calculating the PMR yields 
\eq{
P_{ij}=-\tfrac{4\,\sigma}{\ell}\,E_{ij}^{\SO}\,
}{eq:CG18}
as $\rho_c \to 0$. Like $\tau_{ij}$, the PMR is finite and does not require holographic renormalization.

Given the expressions \eqref{eq:CG17}, \eqref{eq:CG18} for the response functions, the trace conditions \eqref{eq:Trace} follow from tracelessness of the electric and magnetic parts of the Weyl tensor, which give identities 
$\ga^{ij}_{\LO}E_{ij}^{\TO}=\psi_{\FO}^{ij}E_{ij}^{\SO}$ and $\ga^{ij}_{\LO}E_{ij}^{\SO}=\ga^{ij}_{\LO}B_{ijk}^{\FO}=0$. Note that for Starobinsky boundary conditions the Brown-York stress tensor is traceless, but in general only the PMR is traceless.


To summarize, we have shown the consistency of the boundary conditions \eqref{eq:CG6}-\eqref{eq:CG8} by demonstrating that the variational principle is well-defined for the action \eqref{eq:CG11} and by proving finiteness of all 0- and 1-point functions. This is our main result.


Conserved charges may be computed from the currents 
\eq{
 J^{i} = (2\,\tau^{i}{}_{j} + 2\,P^{ik}\,\gamma^{\FO}_{k j})\,\xi^{j} ~,
}{eq:CG20a}
where $\xi^{j}$ is a boundary diffeomorphism associated with an asymptotic symmetry of the theory. (For now we consider only the AdS case $\sigma=-1$, so that the conformal boundary $\partial {\cal M}$ is a timelike surface.) Given a constant-time surface ${\cal C}$ in $\partial {\cal M}$, the charge is 
\eq{
 Q[\xi] = \int_{\cal C}\extd^2x \,\sqrt{h}\,u_{i} J^{i}
}{eq:Charge} 
where $h$ is the metric on ${\cal C}$ and $u^{i}$ is the future-pointing unit vector normal to ${\cal C}$. The combination of $\tau_{ij}$, $P_{ij}$, and $\ga^{\FO}_{ij}$ appearing in $J^{i}$ is precisely the modified stress tensor of Hollands, Ishibashi, and Marolf \cite{Hollands:2005ya}. Thus, the charges \eqref{eq:Charge} are expected to generate the asymptotic symmetries. The covariant divergence of the modified stress tensor satisfies 
\eq{
 {\cal D}_{i} \big(2 \tau^{ij} + 2 P^{i}{}_{k}\,\gamma^{\FO kj}\big) = P^{ik} {\cal D}^{j} \gamma^{\FO}_{ik}  ~.
}{eq:CG19}
This ensures that the difference in charges computed on surfaces ${\cal C}_{1}$, ${\cal C}_{2}$ that bound a region ${\cal V} \subset \partial{\cal M}$ is given by
\eq{
 \Delta Q[\xi] = \int_{\cal V}\extd^3x \sqrt{|\ga^{\LO}|}\,\Big( \tau^{ij} \pounds_{\xi} \ga^{\LO}_{ij} + P^{ij} \pounds_{\xi} \ga^{\FO}_{ij} \Big) ~,
}{eq:lalapetz}
which vanishes for asymptotic symmetries.

We apply now our formulas to three pertinent examples. As a first special case consider solutions that obey Starobinsky boundary conditions, $\ga_{ij}^{\FO}=0$. This includes the asymptotically (A)dS solutions of Einstein gravity with a cosmological constant. Then the EOM imply $E_{ij}^{\SO}=0$, so the PMR vanishes and the Brown--York stress tensor simplifies to
\eq{
\tau_{ij} = \tfrac{4 \sigma}{\ell}\,E_{ij}^{\TO}\,.
}{eq:CG26} 
This recovers the traceless and conserved stress tensor of Einstein gravity \cite{deHaro:2000xn}, in agreement with Maldacena's analysis \cite{Maldacena:2011mk} and with earlier work by Deser and Tekin \cite{Deser:2002rt}. 

A more interesting example is provided by the MKR solution \eqref{eq:CG4}, \eqref{eq:CG5}. 
Setting $\sigma=-1$ for concreteness, and defining the traceless matrix $p^i{}_j=\textrm{diag} (1,\,-\tfrac12,\,-\tfrac12)^i{}_j$ and constants $a_M=(1-\sqrt{1-12aM})/6$ and $m = M/\ell^2$ 
yields
$\tau^i{}_j 
= -8 \frac{m}{\ell^{2}}\,p^i{}_j + 8 \frac{a\,a_M}{\ell^{2}}\,\textrm{diag} (1,\,-1,\,-1)^i{}_j$ 
and
$P^i{}_j 
= 8\,\frac{a_M}{\ell^2}\,p^i{}_j$. 
For vanishing Rindler acceleration, $a=a_M=0$, the previous Einstein case is recovered. For non-vanishing Rindler acceleration, $a\neq0$, the PMR is linear and the trace of the Brown-York stress tensor quadratic in the Rindler parameter $a$ when $aM \ll 1$. Thus, the Rindler parameter in the MKR solution can be interpreted as coming from a partially massless graviton condensate. The conserved charge associated with the Killing vector $\partial_t$ may be computed using \eqref{eq:Charge}. If we normalize the action such that $\alpha_\ts{CG} = \tfrac{1}{64\pi}$ we obtain
  $Q[\partial_t] = m - a\,a_M$.
The entropy, obtained using Wald's approach or from the on-shell action, is
  $S = A_{h}/(4\,\ell^2)$,
where $A_h = 4\pi r_h^{\,2}$ is the area of the horizon $\killi(r_h) = 0$.
Remarkably, the entropy obeys an area law despite the fact that CG is a higher-derivative theory.

As a third example we consider rotating black hole solutions in AdS with Rindler hair parametrized by a Rindler acceleration $\mu$ and rotation parameter $\tilde a$, but with vanishing mass parameter; see Eq.~(7) of \cite{Liu:2012xn}. Interestingly, we find that the absence of a mass parameter leads to a vanishing PMR, $P_{ij}=0$. This shows that a non-zero Rindler term in the asymptotic expansion \eqref{eq:CG7}, $\ga_{ij}^{\FO}\neq 0$, is necessary but not sufficient for a non-zero PMR. Evaluation of the Brown--York stress tensor leads to a conserved energy, $E = -\tilde a^2 \mu/[\ell^2(1-\tilde a^2/\ell^2)^2]$, and conserved angular momentum, $J = E \ell^2/\tilde a$, both linear in the Rindler parameter $\mu$.

Finally, it is possible to make a Legendre transformation of the action \eqref{eq:CG11} that exchanges the role of the PMR and its source, namely by adding a Weyl invariant boundary term 
\eq{
\tilde\Gamma_{\textrm{\tiny CG}} = \Gamma_{\textrm{\tiny CG}} + 8 \int_{\partial{\cal M}}\!\!\!\!\extd^3x\sqrt{|\ga|}\,K^{ij}E_{ij}\,.
}{eq:CG42}
This action is also finite on-shell. Its first variation yields
\eq{
\de\tilde\Gamma_{\textrm{\tiny CG}} = \int_{\partial{\cal M}}\!\!\!\!\extd^3x\sqrt{|\ga|}\,\big(\tilde\tau^{ij}\,\de\ga^{\LO}_{ij}+\tilde{P}^{ij}\,\de E^{\SO}_{ij}\big)
}{eq:CG43}
with finite response functions. 
\begin{align}
\tilde{\tau}_{ij} &= \tau_{ij}+\tfrac{2\sigma}{\ell} \,E_{\SO}^{kl}\psi^{\FO}_{kl}\,\ga^{\LO}_{ij} + \tfrac{8\sigma}{3\ell} \, E^{\SO}_{ij}\ga^{\FO} \nonumber\\
&\quad - \tfrac{4\sigma}{\ell}\,\big(E^{\SO}_{ik}\psi^{\FO k}_j + E^{\SO}_{jk}\psi^{\FO k}_i\big) \\
\tilde{P}_{ij}&=\tfrac{4\sigma}{\ell}\,\ga_{ij}^{\FO} 
\end{align}
The Brown--York stress tensor has zero trace, $\tilde{\tau}^i{}_i=0$. 

To summarize, the results of this Letter provide the basis for CG holography in four dimensions and show the viability of the MKR solution and other solutions with an asymptotic Rindler term.
Possible next steps are the determination of the asymptotic symmetry group, calculation of higher $n$-point functions, and applications of our results to additional solutions of CG.

\acknowledgments

We are grateful to H.~Afshar, S.~Deser, N.~Johansson, A.~Naseh, K.~Skenderis, A.~Waldron and T.~Zojer for discussions.
Many of the calculations presented in this paper were performed with
the {\tt xAct} package for {\it Mathematica} \cite{xAct}.
D.G., M.I., and I.L.~were supported by the START project Y~435-N16 of the Austrian Science Fund (FWF) and the FWF project I~952-N16.


\begin{thebibliography}{44}
\expandafter\ifx\csname natexlab\endcsname\relax\def\natexlab#1{#1}\fi
\expandafter\ifx\csname bibnamefont\endcsname\relax
  \def\bibnamefont#1{#1}\fi
\expandafter\ifx\csname bibfnamefont\endcsname\relax
  \def\bibfnamefont#1{#1}\fi
\expandafter\ifx\csname citenamefont\endcsname\relax
  \def\citenamefont#1{#1}\fi
\expandafter\ifx\csname url\endcsname\relax
  \def\url#1{\texttt{#1}}\fi
\expandafter\ifx\csname urlprefix\endcsname\relax\def\urlprefix{URL }\fi
\providecommand{\bibinfo}[2]{#2}
\providecommand{\eprint}[2][]{\url{#2}}

\bibitem[{\citenamefont{Stelle}(1977)}]{Stelle:1976gc}
\bibinfo{author}{\bibfnamefont{K.}~\bibnamefont{Stelle}},
  \bibinfo{journal}{Phys.Rev.} \textbf{\bibinfo{volume}{D16}},
  \bibinfo{pages}{953} (\bibinfo{year}{1977}).

\bibitem[{\citenamefont{Adler}(1982)}]{Adler:1982ri}
\bibinfo{author}{\bibfnamefont{S.~L.} \bibnamefont{Adler}},
  \bibinfo{journal}{Rev.Mod.Phys.} \textbf{\bibinfo{volume}{54}},
  \bibinfo{pages}{729} (\bibinfo{year}{1982}).

\bibitem[{\citenamefont{Goroff and Sagnotti}(1986)}]{Goroff:1986th}
\bibinfo{author}{\bibfnamefont{M.~H.} \bibnamefont{Goroff}} \bibnamefont{and}
  \bibinfo{author}{\bibfnamefont{A.}~\bibnamefont{Sagnotti}},
  \bibinfo{journal}{Nucl. Phys.} \textbf{\bibinfo{volume}{B266}},
  \bibinfo{pages}{709} (\bibinfo{year}{1986}).

\bibitem[{\citenamefont{Julve and Tonin}(1978)}]{Julve:1978xn}
\bibinfo{author}{\bibfnamefont{J.}~\bibnamefont{Julve}} \bibnamefont{and}
  \bibinfo{author}{\bibfnamefont{M.}~\bibnamefont{Tonin}},
  \bibinfo{journal}{Nuovo Cim.} \textbf{\bibinfo{volume}{B46}},
  \bibinfo{pages}{137} (\bibinfo{year}{1978}).

\bibitem[{\citenamefont{Fradkin and Tseytlin}(1982)}]{Fradkin:1981iu}
\bibinfo{author}{\bibfnamefont{E.}~\bibnamefont{Fradkin}} \bibnamefont{and}
  \bibinfo{author}{\bibfnamefont{A.~A.} \bibnamefont{Tseytlin}},
  \bibinfo{journal}{Nucl.Phys.} \textbf{\bibinfo{volume}{B201}},
  \bibinfo{pages}{469} (\bibinfo{year}{1982}).

\bibitem[{\citenamefont{Tomboulis}(1984)}]{Tomboulis:1983sw}
\bibinfo{author}{\bibfnamefont{E.}~\bibnamefont{Tomboulis}},
  \bibinfo{journal}{Phys.Rev.Lett.} \textbf{\bibinfo{volume}{52}},
  \bibinfo{pages}{1173} (\bibinfo{year}{1984}).

\bibitem[{\citenamefont{Boulware et~al.}(1983)\citenamefont{Boulware, Horowitz,
  and Strominger}}]{Boulware:1983td}
\bibinfo{author}{\bibfnamefont{D.~G.} \bibnamefont{Boulware}},
  \bibinfo{author}{\bibfnamefont{G.~T.} \bibnamefont{Horowitz}},
  \bibnamefont{and}
  \bibinfo{author}{\bibfnamefont{A.}~\bibnamefont{Strominger}},
  \bibinfo{journal}{Phys.Rev.Lett.} \textbf{\bibinfo{volume}{50}},
  \bibinfo{pages}{1726} (\bibinfo{year}{1983}).

\bibitem[{\citenamefont{Mannheim and Kazanas}(1989)}]{Mannheim:1988dj}
\bibinfo{author}{\bibfnamefont{P.~D.} \bibnamefont{Mannheim}} \bibnamefont{and}
  \bibinfo{author}{\bibfnamefont{D.}~\bibnamefont{Kazanas}},
  \bibinfo{journal}{Astrophys. J.} \textbf{\bibinfo{volume}{342}},
  \bibinfo{pages}{635} (\bibinfo{year}{1989}).

\bibitem[{\citenamefont{Mannheim}(2006)}]{Mannheim:2005bfa}
\bibinfo{author}{\bibfnamefont{P.~D.} \bibnamefont{Mannheim}},
  \bibinfo{journal}{Prog. Part. Nucl. Phys.} \textbf{\bibinfo{volume}{56}},
  \bibinfo{pages}{340} (\bibinfo{year}{2006}), \eprint{astro-ph/0505266}.

\bibitem[{\citenamefont{Mannheim and O'Brien}(2011)}]{Mannheim:2010ti}
\bibinfo{author}{\bibfnamefont{P.~D.} \bibnamefont{Mannheim}} \bibnamefont{and}
  \bibinfo{author}{\bibfnamefont{J.~G.} \bibnamefont{O'Brien}},
  \bibinfo{journal}{Phys.Rev.Lett.} \textbf{\bibinfo{volume}{106}},
  \bibinfo{pages}{121101} (\bibinfo{year}{2011}), \eprint{1007.0970}.

\bibitem[{\citenamefont{Mannheim}(2012)}]{Mannheim:2011ds}
\bibinfo{author}{\bibfnamefont{P.~D.} \bibnamefont{Mannheim}},
  \bibinfo{journal}{Found.Phys.} \textbf{\bibinfo{volume}{42}},
  \bibinfo{pages}{388} (\bibinfo{year}{2012}), \eprint{1101.2186}.

\bibitem[{\citenamefont{Berkovits and Witten}(2004)}]{Berkovits:2004jj}
\bibinfo{author}{\bibfnamefont{N.}~\bibnamefont{Berkovits}} \bibnamefont{and}
  \bibinfo{author}{\bibfnamefont{E.}~\bibnamefont{Witten}},
  \bibinfo{journal}{JHEP} \textbf{\bibinfo{volume}{0408}}, \bibinfo{pages}{009}
  (\bibinfo{year}{2004}), \eprint{hep-th/0406051}.

\bibitem[{\citenamefont{Liu and Tseytlin}(1998)}]{Liu:1998bu}
\bibinfo{author}{\bibfnamefont{H.}~\bibnamefont{Liu}} \bibnamefont{and}
  \bibinfo{author}{\bibfnamefont{A.~A.} \bibnamefont{Tseytlin}},
  \bibinfo{journal}{Nucl. Phys.} \textbf{\bibinfo{volume}{B533}},
  \bibinfo{pages}{88} (\bibinfo{year}{1998}), \eprint{hep-th/9804083}.

\bibitem[{\citenamefont{Balasubramanian
  et~al.}(2001)\citenamefont{Balasubramanian, Gimon, Minic, and
  Rahmfeld}}]{Balasubramanian:2000pq}
\bibinfo{author}{\bibfnamefont{V.}~\bibnamefont{Balasubramanian}},
  \bibinfo{author}{\bibfnamefont{E.~G.} \bibnamefont{Gimon}},
  \bibinfo{author}{\bibfnamefont{D.}~\bibnamefont{Minic}}, \bibnamefont{and}
  \bibinfo{author}{\bibfnamefont{J.}~\bibnamefont{Rahmfeld}},
  \bibinfo{journal}{Phys.Rev.} \textbf{\bibinfo{volume}{D63}},
  \bibinfo{pages}{104009} (\bibinfo{year}{2001}), \eprint{hep-th/0007211}.

\bibitem[{\citenamefont{'t~Hooft}(2011)}]{tHooft:2011aa}
\bibinfo{author}{\bibfnamefont{G.}~\bibnamefont{'t~Hooft}},
  \bibinfo{journal}{Found.Phys.} \textbf{\bibinfo{volume}{41}},
  \bibinfo{pages}{1829} (\bibinfo{year}{2011}), \eprint{1104.4543}.

\bibitem[{\citenamefont{Maldacena}(2011)}]{Maldacena:2011mk}
\bibinfo{author}{\bibfnamefont{J.}~\bibnamefont{Maldacena}}
  (\bibinfo{year}{2011}), \eprint{1105.5632}.

\bibitem[{\citenamefont{Starobinsky}(1983)}]{Starobinsky:1982mr}
\bibinfo{author}{\bibfnamefont{A.~A.} \bibnamefont{Starobinsky}},
  \bibinfo{journal}{JETP Lett.} \textbf{\bibinfo{volume}{37}},
  \bibinfo{pages}{66} (\bibinfo{year}{1983}).

\bibitem[{\citenamefont{Anninos et~al.}(2011)\citenamefont{Anninos, Ng, and
  Strominger}}]{Anninos:2010zf}
\bibinfo{author}{\bibfnamefont{D.}~\bibnamefont{Anninos}},
  \bibinfo{author}{\bibfnamefont{G.~S.} \bibnamefont{Ng}}, \bibnamefont{and}
  \bibinfo{author}{\bibfnamefont{A.}~\bibnamefont{Strominger}},
  \bibinfo{journal}{Class.Quant.Grav.} \textbf{\bibinfo{volume}{28}},
  \bibinfo{pages}{175019} (\bibinfo{year}{2011}), \eprint{1009.4730}.

\bibitem[{\citenamefont{Anninos et~al.}(2012)\citenamefont{Anninos, Ng, and
  Strominger}}]{Anninos:2011jp}
\bibinfo{author}{\bibfnamefont{D.}~\bibnamefont{Anninos}},
  \bibinfo{author}{\bibfnamefont{G.~S.} \bibnamefont{Ng}}, \bibnamefont{and}
  \bibinfo{author}{\bibfnamefont{A.}~\bibnamefont{Strominger}},
  \bibinfo{journal}{JHEP} \textbf{\bibinfo{volume}{1202}}, \bibinfo{pages}{032}
  (\bibinfo{year}{2012}), \eprint{1106.1175}.

\bibitem[{\citenamefont{Afshar et~al.}(2011)\citenamefont{Afshar, Cvetkovic,
  Ertl, Grumiller, and Johansson}}]{Afshar:2011yh}
\bibinfo{author}{\bibfnamefont{H.}~\bibnamefont{Afshar}},
  \bibinfo{author}{\bibfnamefont{B.}~\bibnamefont{Cvetkovic}},
  \bibinfo{author}{\bibfnamefont{S.}~\bibnamefont{Ertl}},
  \bibinfo{author}{\bibfnamefont{D.}~\bibnamefont{Grumiller}},
  \bibnamefont{and}
  \bibinfo{author}{\bibfnamefont{N.}~\bibnamefont{Johansson}},
  \bibinfo{journal}{Phys.Rev.} \textbf{\bibinfo{volume}{D84}},
  \bibinfo{pages}{041502(R)} (\bibinfo{year}{2011}), \eprint{1106.6299}.
%
  \bibinfo{journal}{Phys.Rev.} \textbf{\bibinfo{volume}{D85}},
  \bibinfo{pages}{064033} (\bibinfo{year}{2012}), \eprint{1110.5644}.

\bibitem[{\citenamefont{Riegert}(1984)}]{Riegert:1984zz}
\bibinfo{author}{\bibfnamefont{R.~J.} \bibnamefont{Riegert}},
  \bibinfo{journal}{Phys.Rev.Lett.} \textbf{\bibinfo{volume}{53}},
  \bibinfo{pages}{315} (\bibinfo{year}{1984}).

\bibitem[{\citenamefont{Grumiller}(2010)}]{Grumiller:2010bz}
\bibinfo{author}{\bibfnamefont{D.}~\bibnamefont{Grumiller}},
  \bibinfo{journal}{Phys.Rev.Lett.} \textbf{\bibinfo{volume}{105}},
  \bibinfo{pages}{211303} (\bibinfo{year}{2010}), \eprint{1011.3625}.

\bibitem[{\citenamefont{York}(1972)}]{York:1972sj}
\bibinfo{author}{\bibfnamefont{J.~W.} \bibnamefont{York}, \bibfnamefont{Jr.}},
  \bibinfo{journal}{Phys. Rev. Lett.} \textbf{\bibinfo{volume}{28}},
  \bibinfo{pages}{1082} (\bibinfo{year}{1972}).

\bibitem[{\citenamefont{Gibbons and Hawking}(1977)}]{Gibbons:1977ue}
\bibinfo{author}{\bibfnamefont{G.~W.} \bibnamefont{Gibbons}} \bibnamefont{and}
  \bibinfo{author}{\bibfnamefont{S.~W.} \bibnamefont{Hawking}},
  \bibinfo{journal}{Phys. Rev.} \textbf{\bibinfo{volume}{D15}},
  \bibinfo{pages}{2752} (\bibinfo{year}{1977}).

\bibitem[{\citenamefont{Henningson and Skenderis}(2000)}]{Henningson:1998ey}
\bibinfo{author}{\bibfnamefont{M.}~\bibnamefont{Henningson}} \bibnamefont{and}
  \bibinfo{author}{\bibfnamefont{K.}~\bibnamefont{Skenderis}},
  \bibinfo{journal}{Fortsch. Phys.} \textbf{\bibinfo{volume}{48}},
  \bibinfo{pages}{125} (\bibinfo{year}{2000}), \eprint{hep-th/9812032}.

\bibitem[{\citenamefont{Balasubramanian and
  Kraus}(1999)}]{Balasubramanian:1999re}
\bibinfo{author}{\bibfnamefont{V.}~\bibnamefont{Balasubramanian}}
  \bibnamefont{and} \bibinfo{author}{\bibfnamefont{P.}~\bibnamefont{Kraus}},
  \bibinfo{journal}{Commun. Math. Phys.} \textbf{\bibinfo{volume}{208}},
  \bibinfo{pages}{413} (\bibinfo{year}{1999}), \eprint{hep-th/9902121}.

\bibitem[{\citenamefont{Emparan et~al.}(1999)\citenamefont{Emparan, Johnson,
  and Myers}}]{Emparan:1999pm}
\bibinfo{author}{\bibfnamefont{R.}~\bibnamefont{Emparan}},
  \bibinfo{author}{\bibfnamefont{C.~V.} \bibnamefont{Johnson}},
  \bibnamefont{and} \bibinfo{author}{\bibfnamefont{R.~C.} \bibnamefont{Myers}},
  \bibinfo{journal}{Phys. Rev.} \textbf{\bibinfo{volume}{D60}},
  \bibinfo{pages}{104001} (\bibinfo{year}{1999}), \eprint{hep-th/9903238}.

\bibitem[{\citenamefont{Kraus et~al.}(1999)\citenamefont{Kraus, Larsen, and
  Siebelink}}]{Kraus:1999di}
\bibinfo{author}{\bibfnamefont{P.}~\bibnamefont{Kraus}},
  \bibinfo{author}{\bibfnamefont{F.}~\bibnamefont{Larsen}}, \bibnamefont{and}
  \bibinfo{author}{\bibfnamefont{R.}~\bibnamefont{Siebelink}},
  \bibinfo{journal}{Nucl. Phys.} \textbf{\bibinfo{volume}{B563}},
  \bibinfo{pages}{259} (\bibinfo{year}{1999}), \eprint{hep-th/9906127}.

\bibitem[{\citenamefont{de~Haro et~al.}(2001)\citenamefont{de~Haro, Solodukhin,
  and Skenderis}}]{deHaro:2000xn}
\bibinfo{author}{\bibfnamefont{S.}~\bibnamefont{de~Haro}},
  \bibinfo{author}{\bibfnamefont{S.~N.} \bibnamefont{Solodukhin}},
  \bibnamefont{and}
  \bibinfo{author}{\bibfnamefont{K.}~\bibnamefont{Skenderis}},
  \bibinfo{journal}{Commun. Math. Phys.} \textbf{\bibinfo{volume}{217}},
  \bibinfo{pages}{595} (\bibinfo{year}{2001}), \eprint{hep-th/0002230}.

\bibitem[{\citenamefont{Papadimitriou and
  Skenderis}(2005)}]{Papadimitriou:2005ii}
\bibinfo{author}{\bibfnamefont{I.}~\bibnamefont{Papadimitriou}}
  \bibnamefont{and}
  \bibinfo{author}{\bibfnamefont{K.}~\bibnamefont{Skenderis}},
  \bibinfo{journal}{JHEP} \textbf{\bibinfo{volume}{08}}, \bibinfo{pages}{004}
  (\bibinfo{year}{2005}), \eprint{hep-th/0505190}.

\bibitem[{\citenamefont{Lu et~al.}(2012)\citenamefont{Lu, Pang, Pope, and
  Vazquez-Poritz}}]{Lu:2012xu}
\bibinfo{author}{\bibfnamefont{H.}~\bibnamefont{Lu}},
  \bibinfo{author}{\bibfnamefont{Y.}~\bibnamefont{Pang}},
  \bibinfo{author}{\bibfnamefont{C.}~\bibnamefont{Pope}}, \bibnamefont{and}
  \bibinfo{author}{\bibfnamefont{J.~F.} \bibnamefont{Vazquez-Poritz}},
  \bibinfo{journal}{Phys.Rev.} \textbf{\bibinfo{volume}{D86}},
  \bibinfo{pages}{044011} (\bibinfo{year}{2012}), \eprint{1204.1062}.

\bibitem[{\citenamefont{Wald}(1993)}]{Wald:1993nt}
\bibinfo{author}{\bibfnamefont{R.~M.} \bibnamefont{Wald}},
  \bibinfo{journal}{Phys. Rev.} \textbf{\bibinfo{volume}{D48}},
  \bibinfo{pages}{3427} (\bibinfo{year}{1993}),
  \eprint[http://arXiv.org/abs]{gr-qc/9307038}.

\bibitem[{\citenamefont{Myers}(1987)}]{Myers:1987yn}
\bibinfo{author}{\bibfnamefont{R.~C.} \bibnamefont{Myers}},
  \bibinfo{journal}{Phys. Rev.} \textbf{\bibinfo{volume}{D36}},
  \bibinfo{pages}{392} (\bibinfo{year}{1987}).

\bibitem[{\citenamefont{Deser and Nepomechie}(1984)}]{Deser:1983mm}
\bibinfo{author}{\bibfnamefont{S.}~\bibnamefont{Deser}} \bibnamefont{and}
  \bibinfo{author}{\bibfnamefont{R.~I.} \bibnamefont{Nepomechie}},
  \bibinfo{journal}{Ann. Phys.} \textbf{\bibinfo{volume}{154}},
  \bibinfo{pages}{396} (\bibinfo{year}{1984}).

\bibitem[{\citenamefont{Deser and Waldron}(2001)}]{Deser:2001pe}
\bibinfo{author}{\bibfnamefont{S.}~\bibnamefont{Deser}} \bibnamefont{and}
  \bibinfo{author}{\bibfnamefont{A.}~\bibnamefont{Waldron}},
  \bibinfo{journal}{Phys. Rev. Lett.} \textbf{\bibinfo{volume}{87}},
  \bibinfo{pages}{031601} (\bibinfo{year}{2001}), \eprint{hep-th/0102166}.

\bibitem[{\citenamefont{Deser et~al.}(2012)\citenamefont{Deser, Joung, and
  Waldron}}]{Deser:2013bs}
\bibinfo{author}{\bibfnamefont{S.}~\bibnamefont{Deser}},
  \bibinfo{author}{\bibfnamefont{E.}~\bibnamefont{Joung}}, \bibnamefont{and}
  \bibinfo{author}{\bibfnamefont{A.}~\bibnamefont{Waldron}},
  \bibinfo{journal}{Phys.Rev.} \textbf{\bibinfo{volume}{D86}},
  \bibinfo{pages}{104004} (\bibinfo{year}{2012}), \eprint{1301.4181}.

\bibitem[{\citenamefont{Deser et~al.}(2013{\natexlab{a}})\citenamefont{Deser,
  Sandora, and Waldron}}]{Deser:2013uy}
\bibinfo{author}{\bibfnamefont{S.}~\bibnamefont{Deser}},
  \bibinfo{author}{\bibfnamefont{M.}~\bibnamefont{Sandora}}, \bibnamefont{and}
  \bibinfo{author}{\bibfnamefont{A.}~\bibnamefont{Waldron}},
  \bibinfo{journal}{Phys.Rev.} \textbf{\bibinfo{volume}{D87}},
  \bibinfo{pages}{101501} (\bibinfo{year}{2013}{\natexlab{a}}),
  \eprint{1301.5621}.

\bibitem[{\citenamefont{Deser et~al.}(2013{\natexlab{b}})\citenamefont{Deser,
  Ertl, and Grumiller}}]{Deser:2012ci}
\bibinfo{author}{\bibfnamefont{S.}~\bibnamefont{Deser}},
  \bibinfo{author}{\bibfnamefont{S.}~\bibnamefont{Ertl}}, \bibnamefont{and}
  \bibinfo{author}{\bibfnamefont{D.}~\bibnamefont{Grumiller}},
  \bibinfo{journal}{J.Phys.} \textbf{\bibinfo{volume}{A46}},
  \bibinfo{pages}{214018} (\bibinfo{year}{2013}{\natexlab{b}}),
  \eprint{1208.0339}.

\bibitem[{\citenamefont{Hollands et~al.}(2005)\citenamefont{Hollands,
  Ishibashi, and Marolf}}]{Hollands:2005ya}
\bibinfo{author}{\bibfnamefont{S.}~\bibnamefont{Hollands}},
  \bibinfo{author}{\bibfnamefont{A.}~\bibnamefont{Ishibashi}},
  \bibnamefont{and} \bibinfo{author}{\bibfnamefont{D.}~\bibnamefont{Marolf}},
  \bibinfo{journal}{Phys.Rev.} \textbf{\bibinfo{volume}{D72}},
  \bibinfo{pages}{104025} (\bibinfo{year}{2005}), \eprint{hep-th/0503105}.

\bibitem[{\citenamefont{Deser and Tekin}(2002)}]{Deser:2002rt}
\bibinfo{author}{\bibfnamefont{S.}~\bibnamefont{Deser}} \bibnamefont{and}
  \bibinfo{author}{\bibfnamefont{B.}~\bibnamefont{Tekin}},
  \bibinfo{journal}{Phys.Rev.Lett.} \textbf{\bibinfo{volume}{89}},
  \bibinfo{pages}{101101} (\bibinfo{year}{2002}), \eprint{hep-th/0205318}.
%
  \bibinfo{journal}{Phys.Rev.} \textbf{\bibinfo{volume}{D67}},
  \bibinfo{pages}{084009} (\bibinfo{year}{2003}), \eprint{hep-th/0212292}.


\bibitem[{\citenamefont{Liu and Lu}(2013)}]{Liu:2012xn}
\bibinfo{author}{\bibfnamefont{H.-S.} \bibnamefont{Liu}} \bibnamefont{and}
  \bibinfo{author}{\bibfnamefont{H.}~\bibnamefont{Lu}}, \bibinfo{journal}{JHEP}
  \textbf{\bibinfo{volume}{1302}}, \bibinfo{pages}{139} (\bibinfo{year}{2013}),
  \eprint{1212.6264}.


\bibitem[{\citenamefont{Bender and Mannheim}(2008)}]{Bender:2007wu}
\bibinfo{author}{\bibfnamefont{C.~M.} \bibnamefont{Bender}} \bibnamefont{and}
  \bibinfo{author}{\bibfnamefont{P.~D.} \bibnamefont{Mannheim}},
  \bibinfo{journal}{Phys.Rev.Lett.} \textbf{\bibinfo{volume}{100}},
  \bibinfo{pages}{110402} (\bibinfo{year}{2008}), \eprint{0706.0207}.

\bibitem[{\citenamefont{Smilga}(2009)}]{Smilga:2008pr}
\bibinfo{author}{\bibfnamefont{A.}~\bibnamefont{Smilga}},
  \bibinfo{journal}{SIGMA} \textbf{\bibinfo{volume}{5}}, \bibinfo{pages}{017}
  (\bibinfo{year}{2009}), \eprint{0808.0139}.

\bibitem[{\citenamefont{Chen et~al.}(2013)\citenamefont{Chen, Fasiello, Lim,
  and Tolley}}]{Chen:2012au}
\bibinfo{author}{\bibfnamefont{T.-j.} \bibnamefont{Chen}},
  \bibinfo{author}{\bibfnamefont{M.}~\bibnamefont{Fasiello}},
  \bibinfo{author}{\bibfnamefont{E.~A.} \bibnamefont{Lim}}, \bibnamefont{and}
  \bibinfo{author}{\bibfnamefont{A.~J.} \bibnamefont{Tolley}},
  \bibinfo{journal}{JCAP} \textbf{\bibinfo{volume}{1302}}, \bibinfo{pages}{042}
  (\bibinfo{year}{2013}), \eprint{1209.0583}.


\bibitem[{\citenamefont{Johansson et~al.}(2012)\citenamefont{Johansson, Naseh,
  and Zojer}}]{Johansson:2012fs}
\bibinfo{author}{\bibfnamefont{N.}~\bibnamefont{Johansson}},
  \bibinfo{author}{\bibfnamefont{A.}~\bibnamefont{Naseh}}, \bibnamefont{and}
  \bibinfo{author}{\bibfnamefont{T.}~\bibnamefont{Zojer}},
  \bibinfo{journal}{JHEP} \textbf{\bibinfo{volume}{1209}}, \bibinfo{pages}{114}
  (\bibinfo{year}{2012}), \eprint{1205.5804}.

\bibitem[{\citenamefont{Lu and Pope}(2011)}]{Lu:2011zk}
\bibinfo{author}{\bibfnamefont{H.}~\bibnamefont{Lu}} \bibnamefont{and}
  \bibinfo{author}{\bibfnamefont{C.}~\bibnamefont{Pope}},
  \bibinfo{journal}{Phys.Rev.Lett.} \textbf{\bibinfo{volume}{106}},
  \bibinfo{pages}{181302} (\bibinfo{year}{2011}), \eprint{1101.1971}.

\bibitem[{\citenamefont{Hohm and Tonni}(2010)}]{Hohm:2010jc}
\bibinfo{author}{\bibfnamefont{O.}~\bibnamefont{Hohm}} \bibnamefont{and}
  \bibinfo{author}{\bibfnamefont{E.}~\bibnamefont{Tonni}},
  \bibinfo{journal}{JHEP} \textbf{\bibinfo{volume}{04}}, \bibinfo{pages}{093}
  (\bibinfo{year}{2010}), \eprint{1001.3598}.

\bibitem[{\citenamefont{Martin-Garcia}(2013)}]{xAct}
\bibinfo{author}{\bibfnamefont{J.M.}~\bibnamefont{Martin-Garcia}},
  \bibinfo{title}{xAct: tensor computer algebra}
 \bibinfo{url}{http://www.xact.es/}.

\end{thebibliography}

\end{document}